\documentclass{appolb}

\usepackage{fleqn,epsf,graphicx}
\usepackage{authblk}
\usepackage{cite}

\usepackage{epsfig}
\usepackage{graphicx}
\usepackage{caption}
\usepackage{subcaption}
\usepackage{hyperref}

\usepackage{amsmath}
\usepackage{xcolor}

\begin{document}
\title{Scatter fraction of the J-PET tomography scanner
}


\author{

P.~Kowalski$^{a}$, W.~Wi\'slicki$^{a}$, L.~Raczy\'nski$^{a}$, D.~Alfs$^{b}$, T.~Bednarski$^{b}$, P.~Bia\l as$^{b}$, E.~Czerwi\'nski$^{b}$, 
A.~Gajos$^{b}$, 
B.~G\l owacz$^{b}$,
J.~Jasi\'nska$^{c}$, D.~Kami\'nska$^{b}$, 
G.~Korcyl$^{b}$, 
T.~Kozik$^{b}$, W.~Krzemie\'n$^{d}$, E.~Kubicz$^{b}$, 
M.~Mohammad$^{b}$, 
Sz.~Nied\'zwiecki$^{b}$, 
M.~Pa\l ka$^{b}$, 
M.~Pawlik-Nied\'zwiecka$^{b}$, 
Z.~Rudy$^{b}$, 
M.~Silarski$^{b}$, 
A.~Wieczorek$^{b,e}$, B.~Zgardzi\'nska$^{c}$, M.~Zieli\'nski$^{b}$, 
P.~Moskal$^{b}$

\address{
$^{a}$ Department of Complex Systems, National Centre for Nuclear Research, 05-400 Otwock-\'Swierk, Poland \\
$^{b}$ Faculty of Physics, Astronomy and Applied Computer Science, Jagiellonian University, 30-348 Cracow, Poland \\
$^{c}$ Faculty of Mathematics, Physics and Computer Science, Maria Curie-Sk\l odowska University, 20-031 Lublin, Poland \\ 
$^{d}$ High Energy Physics Division, National Centre for Nuclear Research, 05-400 Otwock-\'Swierk, Poland \\
$^{e}$ Intitute of Metallurgy and Material Science of Polish Academy of Sciences, 30-059 Cracow, Poland \\
}

}

\maketitle
\begin{abstract}

A novel Positron Emission Tomography system, based on plastic scintillators, is being developed by the J-PET collaboration. 
In this article we present the simulation results of the scatter fraction, representing one of the parameters crucial for background studies defined in the NEMA-NU-2-2012 norm. We elaborate an event selection methods allowing to suppress events in which gamma quanta were scattered in the phantom or underwent  the multiple scattering in the detector.
The estimated scatter fraction for the single-layer J-PET scanner varies from 37\% to 53\% depending on the applied energy threshold.
\end{abstract}
  
\section{Introduction}

A novel prototype PET scanner is being developed by the J-PET collaboration at the Jagiellonian University \cite{jpet_paper_nima_1,jpet_paper_nima_2,jpet_paper_nima_3,
jpet_paper_nima_4,jpet_paper_actaa_1,jpet_paper_actaa_2, jpet_paper_actaa_3,jpet_paper_nukleonika_1, jpet_paper_nukleonika_2}.
Its innovation arises from applying plastic scintillators \cite{jpet_patent_1,jpet_patent_2} instead of inorganic crystals used in classical PET scanners. The rectangular plastic scintillator strips are arranged axially, and form a~cylindrical scanner. The high geometrical acceptance (large field of view) makes this scanner suitable for the simultaneous full-body scans.
Also, the usage of plastic scintillators with its very good time properties allows to apply effectively the time of flight (TOF) technique, which improves the contrast of the reconstructed image.
In addition, the  plastic scintillator material is much cheaper compared to the inorganic crystals used in the current tomography scanners.

In this article we present the results of the scatter fraction determination. The scatter fraction (SF) is one of the parameters defined by the National Electrical Manufacturers Association (NEMA)~\cite{norm_nema}, which characterizes the tomography scanner performance. 
The determination of SF for the J-PET tomograph has been done by Monte-Carlo simulation method taking into account known properties of the detection modules \cite{jpet_paper_nima_1, jpet_paper_nima_2, jpet_paper_nima_3, jpet_paper_nima_4}. These simulations were performed using the \textit{Geant4 Application for Tomographic Emission} (GATE) software \cite{gate_paper1,gate_paper2} based on the GEANT4 toolkit \cite{geant4_paper_1}.

In simulation a~simplified version of the J-PET laboratory prototype, built of a~single-layer cylinder, was used.
Two classes of scattered events have been taken into account:  gamma quanta scattered in the phantom, as well as events in which multiple scattering occurs in the detector.
The analysis does not include the impact of the accidental coincidences, which is described in the reference~\cite{jpet_paper_Kowalski2}.


\section{Scatter fraction}

\begin{figure}[h!]
\centering

\begin{subfigure}{0.325\textwidth}
\centering
\includegraphics[width=\textwidth]{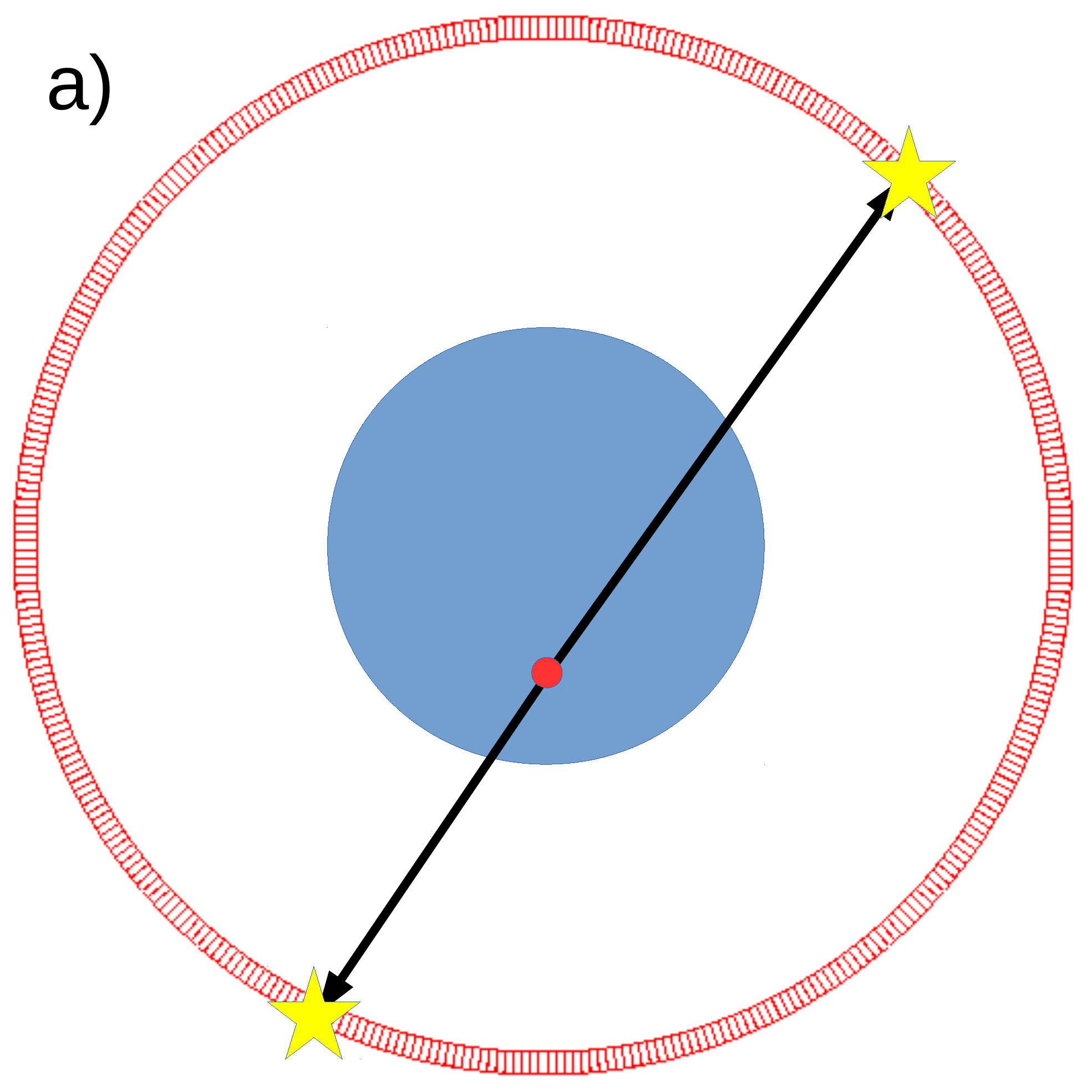}
\end{subfigure}
\begin{subfigure}{0.325\textwidth}
\centering
\includegraphics[width=\textwidth]{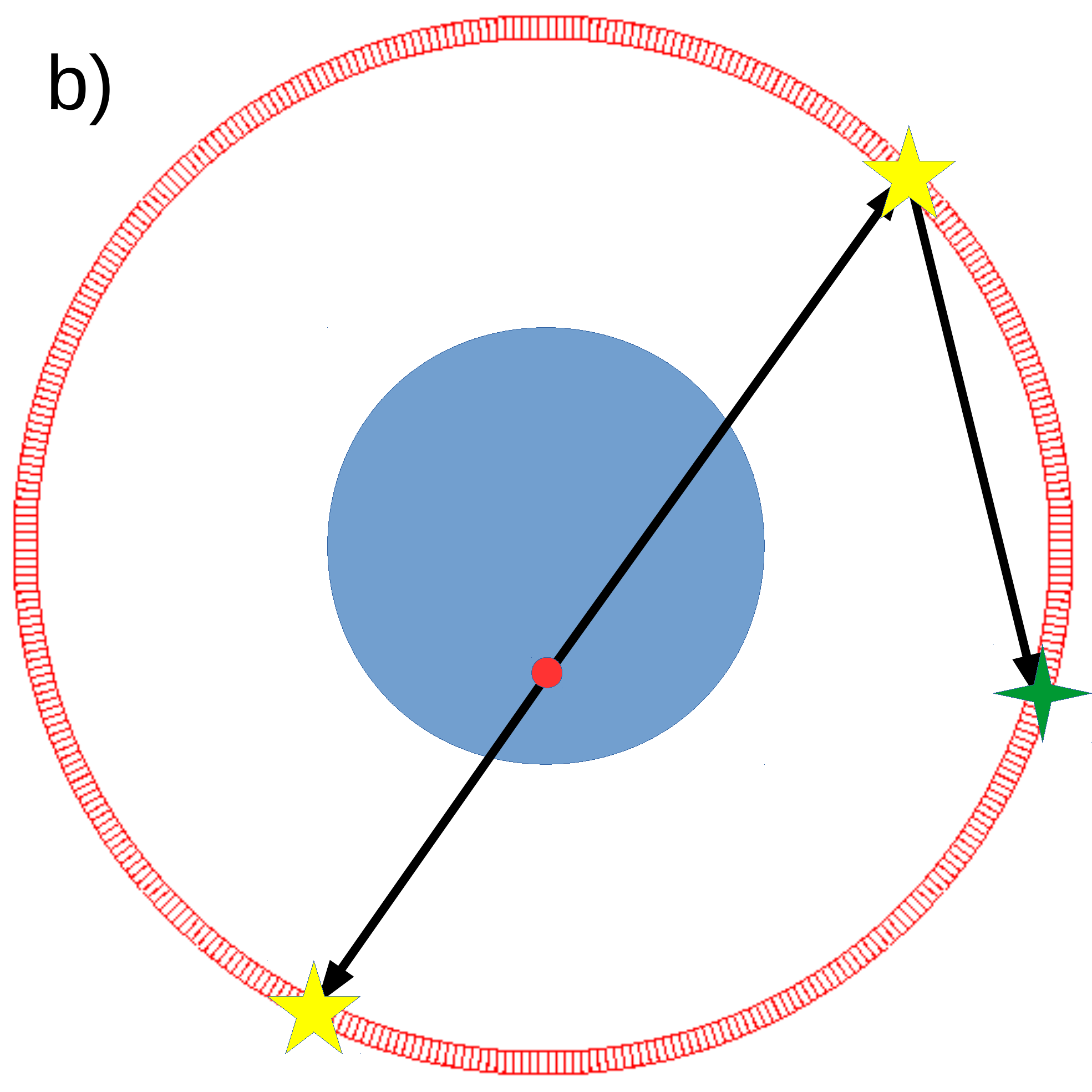}
\end{subfigure}

\begin{subfigure}{0.325\textwidth}
\centering
\includegraphics[width=\textwidth]{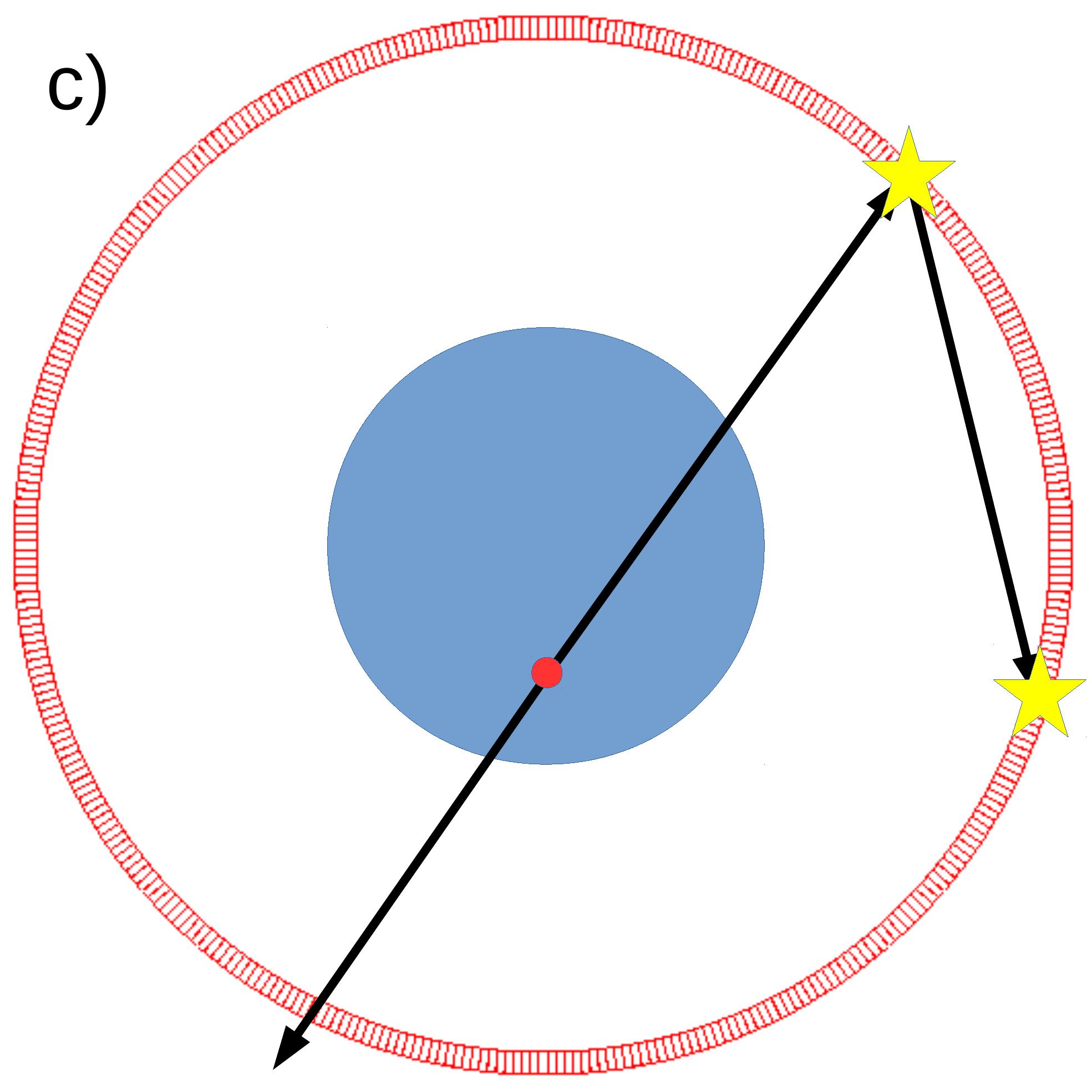}
\end{subfigure}
\begin{subfigure}{0.325\textwidth}
\centering
\includegraphics[width=\textwidth]{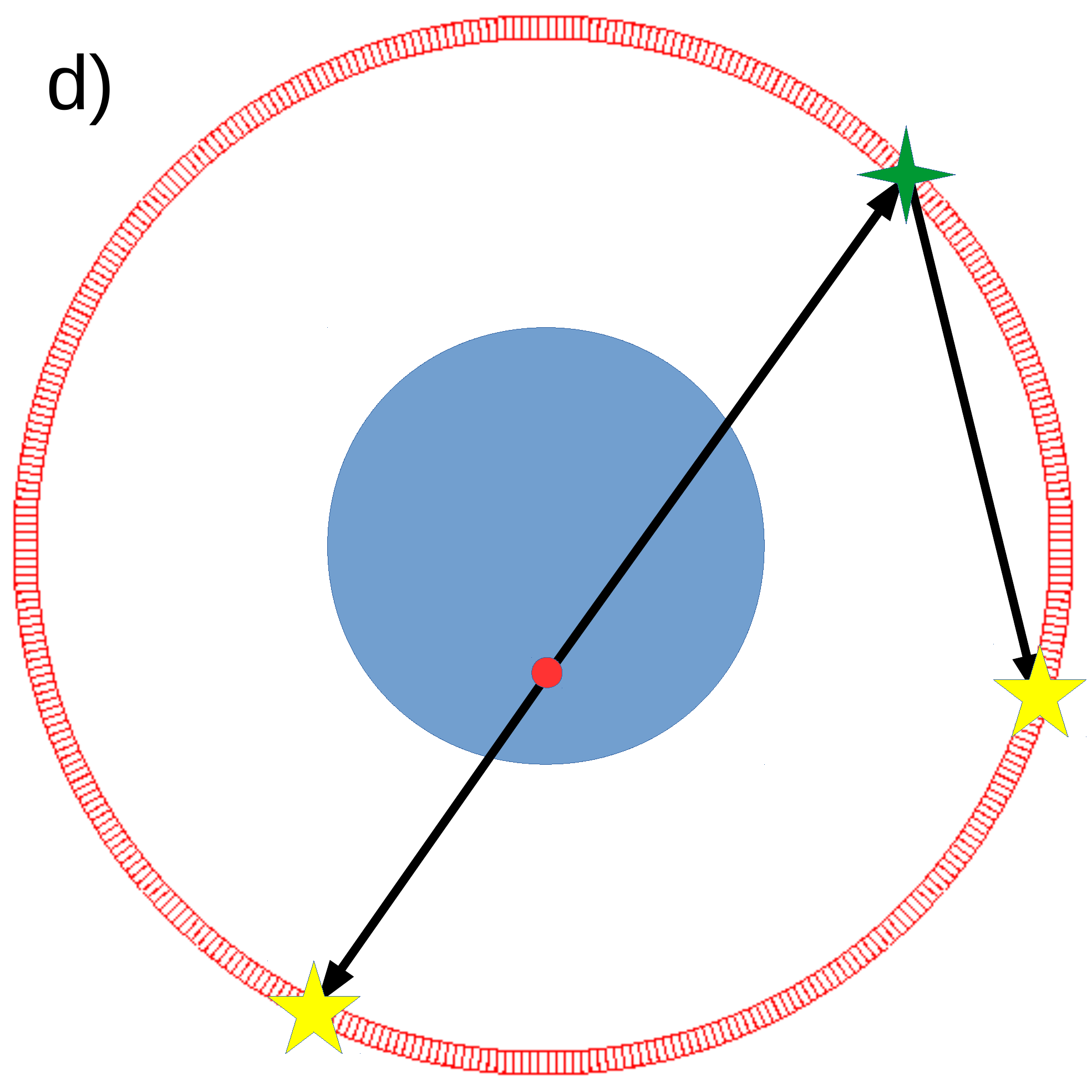}
\end{subfigure}

\begin{subfigure}{0.325\textwidth}
\centering
\includegraphics[width=\textwidth]{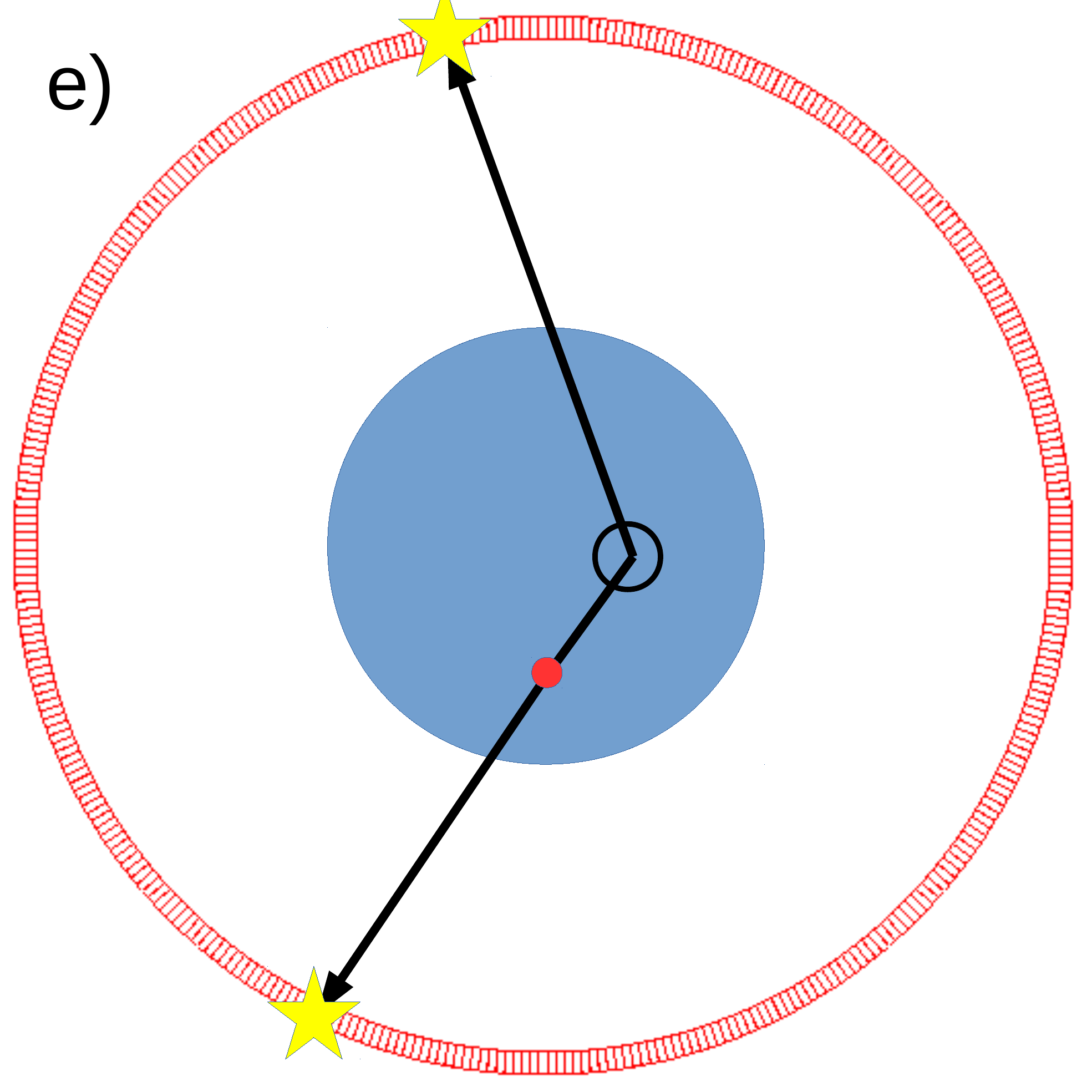}
\end{subfigure}
\begin{subfigure}{0.325\textwidth}
\centering
\includegraphics[width=\textwidth]{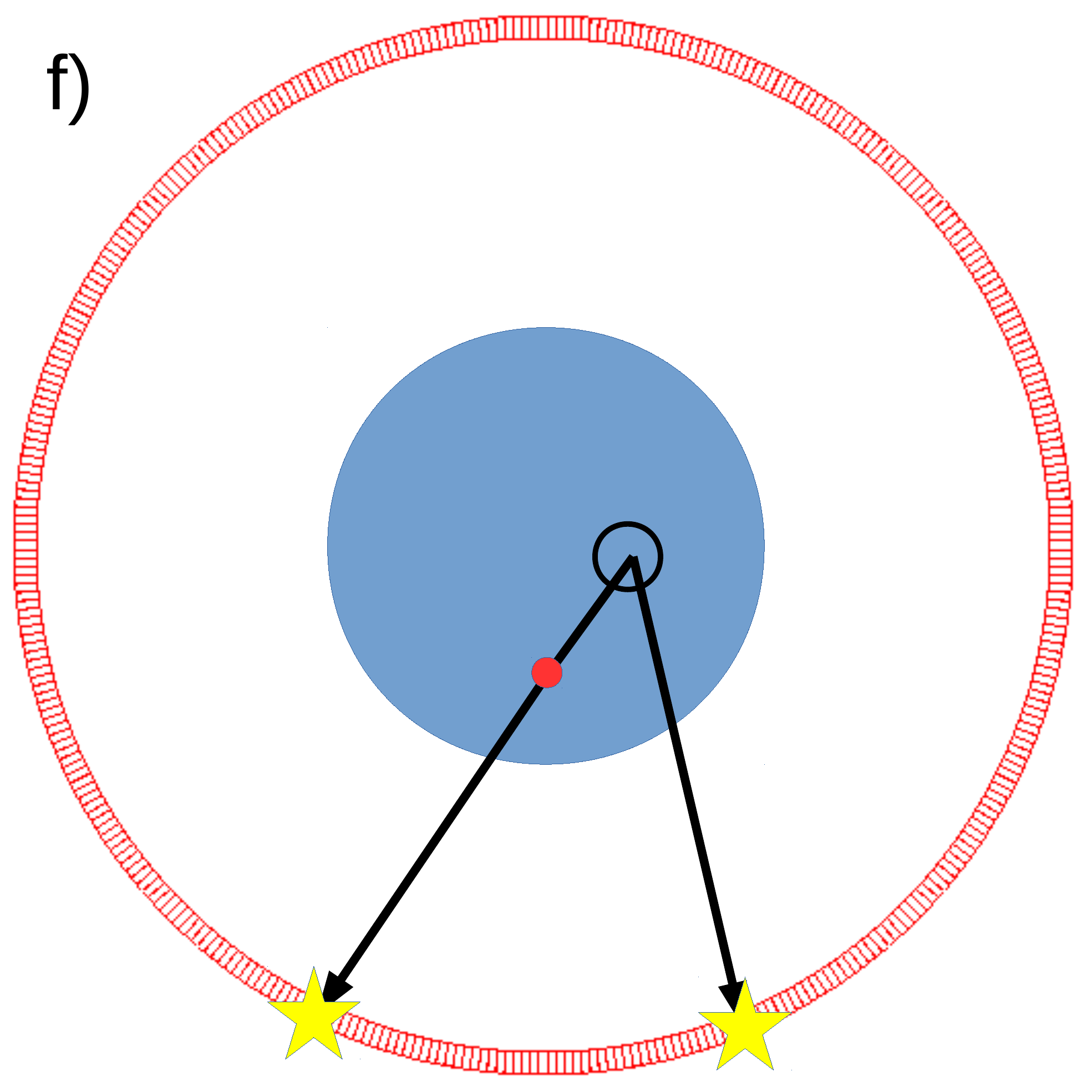}
\end{subfigure}

\begin{subfigure}{0.325\textwidth}
\centering
\includegraphics[width=\textwidth]{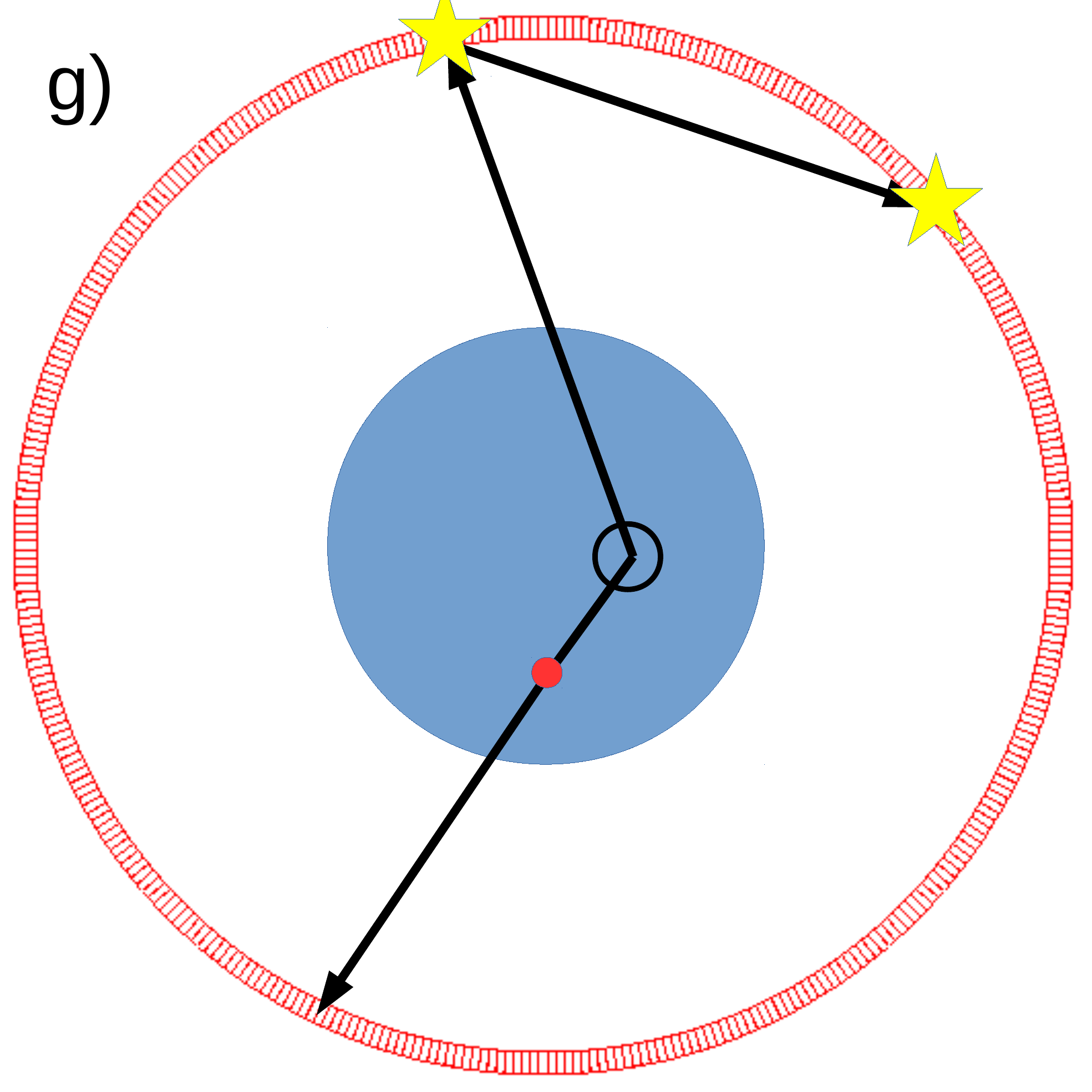}
\end{subfigure}
\begin{subfigure}{0.325\textwidth}
\centering
\includegraphics[width=\textwidth]{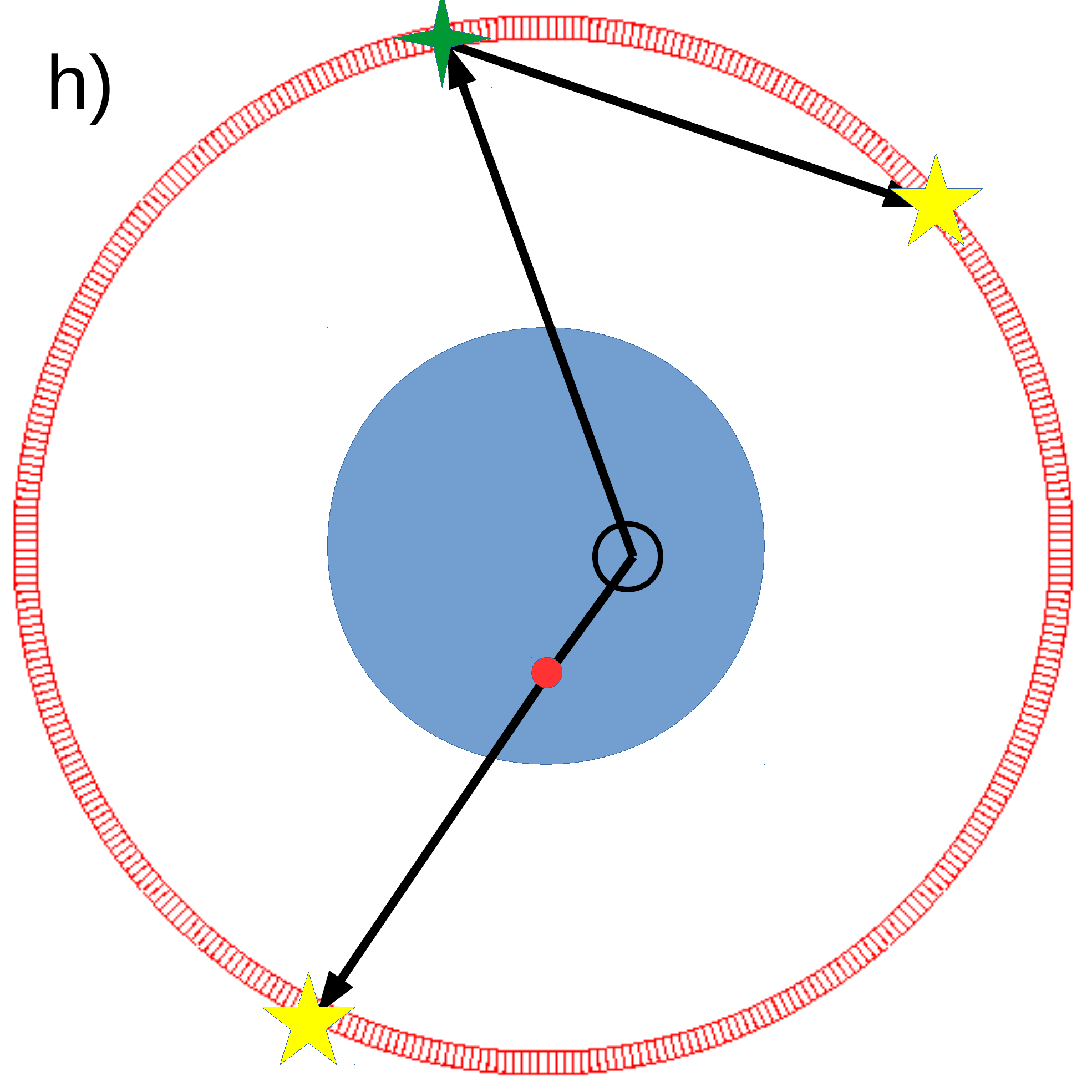}
\end{subfigure}

\caption{Pictorial definitions of different types of coincidences: a) and b) true coincidences, c) and d) detector-scattered coincidence, e) to h) phantom-scattered coincidences. Source is marked as a red filled circle. Places of Compton scatterings with energy deposition bigger than defined energy threshold are marked with five-pointed yellow stars, while scatterings with energy higher than the noise level but lower than the  energy threshold are marked with four-pointed green stars. Place of Compton scattering in the phantom in cases e) to h) is marked with an empty circle.}
\label{types_of_coincidences}
\end{figure}

PET tomography involves registering of 511~keV gamma photons outgoing from the patient's body, where they were created in the positron-electron annihilations. In the plastic scintillators, such a~photon interacts in practice only via Compton interaction with the negligible probability of photoproduction and Rayleigh processes. If two gamma photons are scattered in the detector and each deposits energy larger than the noise threshold (and thus they are registered) in two different scintillators in a fixed time window, such an event is commonly referred to as a coincidence. Certainly a photon may scatter more than once. An example of various events categories are shown in Fig.~\ref{types_of_coincidences}.
Apart from the minimum photon deposited energy in one scattering, for which the photon is treated as registered (so called noise level), in the simulation we also consider fixed energy threshold, which corresponds to the energy level applied in the data analysis.
To avoid the confusion, in the following we will refer to those levels, as noise threshold and fixed energy threshold, respectively. 
The five-pointed yellow stars indicate signals larger than an energy threshold used in the offline analysis, whereas the four-pointed green stars indicate signals exceeding noise level but lower than the energy threshold used. 
In this article we consider only events fulfilling the condition that for exactly two interactions the deposited energy is bigger than the fixed energy threshold and that there is no more than one additional scattering above the noise threshold but below the fixed energy threshold.

Coincidence events fall into three different categories: true, scattered and accidental. Fig. \ref{types_of_coincidences} shows examples of typical events. 
For the image reconstruction only these events are taken into account for which two signals above energy threshold were registered in two different detection modules. 
As it is shown in Fig. \ref{types_of_coincidences} (e.g, c to h)  this definition does not exclude many wrongly reconstructed Line-of-Response (LOR).
True coincidences are shown in Fig. \ref{types_of_coincidences}a and \ref{types_of_coincidences}b. A~true coincidence occurs when both signals, which are above the fixed energy threshold, correspond to the primary interactions of the gamma photons, which means that these photons did not scatter before, neither in the phantom nor in the other detection modules.
Scattered coincidences represent events in which at least one of photons scattered in the detector (detector-scattered coincidences - Fig.~\ref{types_of_coincidences}c,d) or in the phantom (phantom-scattered coincidences - Fig.~\ref{types_of_coincidences}e-h) prior to the interaction with the energy deposition  larger than the fixed energy threshold. In accidental coincidences gamma photons come from different annihilations. As it was aforementioned, the accidental coincidences are not considered in the current simulations. 

The SF parameter is a~dimensionless ratio between the number of scattered coincidences and the sum of scattered and true coincidences in the field of view of the scanner. The smaller the scatter fraction, the better the quality of the reconstructed image \cite{norm_nema}.

Since in case of the J-PET scanner, both phantom- and detector-scattered coincidences occur, we can define the SF in two different ways. First definition (which comes from NEMA-NU-2 norm) includes only phantom-scattered coincidences (Eq. \ref{eq:sf1}). The second definition contains terms corresponding to both types of scattered coincidences (Eq. \ref{eq:sf2}).

\begin{equation} 
\label{eq:sf1}
SF_1 = {N_{psca} \over {N_{true} + N_{psca}}},
\end{equation}

\begin{equation} 
\label{eq:sf2}
SF_2 = {{N_{psca}+N_{dsca}} \over {N_{true} + N_{psca} + N_{dsca}}},
\end{equation}

\noindent where $N_{true}$ is the number of true coincidences, $N_{psca}$ is the number of phantom-scattered coincidences and $N_{dsca}$ is the number of detector-scattered coincidences.

\section{Simulation details}

Simulations were performed using GATE software (version 7.0) \cite{gate_paper1,gate_paper2} at the \'Swierk Computing Centre at National Centre for Nuclear Research.

The J-PET detector was defined as a~cylinder with inner radius R~=~427.8~mm, constructed from the plastic scintillator strips. The detector consisted of 384 rectangular EJ-230 scintillators with dimensions of 7~mm~x~19~mm~x~500~mm. Photomultipliers Hammamatsu R4998 were attached to both ends of each strip. Material properties of EJ-230 scintillator and properties of photomulitplier R4998 were implemented like in the previous work \cite{jpet_paper_Kowalski2}. Simulations were performed under assumption that the detecting chamber, phantom and source are placed in the vacuum, what ensures that there are no scatterings of gamma photons in the space surrounding the scanner.

The simulated phantom and the source were chosen according to the NEMA-NU-2-2012 norm \cite{norm_nema}. The phantom is defined as a solid cylinder composed of polyethylene with a density equal to $0.96 {g \over cm^3}$, the outside diameter 203~mm and the length 700~mm.
Along the cylinder (parallel to the central axis) at a~radial distance of 45~mm, a~hole with diameter 6.4~mm was drilled. In this hole the linear source (polyethylene tube filled with known activity) was placed. Its length was equal to the length of the cylinder and its diameter was 4.8~mm.


According to the NEMA norm, data collected during measurements should contain at least 500.000 prompt counts. Prompt counts are all coincidences detected by the system (true, scattered and accidental). The simulations reported in this article were performed with the activity of the source set to 1~MBq and the simulated time to 100~s.
This resulted in the number of collected events exceeding the required 500.000. The annihilation points were distributed homogeneously within the source and for each event two gamma photons with energy of 511~keV were emitted back-to-back with the isotropic angular distribution. 

\section{Event classification} \label{methods}

The post-simulation analysis was performed using the software developed by the authors. Its main goal was to distinguish between different types of coincidences and to visualize the obtained results. The method to disentangle different types of coincidences was based on the information on number of signals recorded above the noise level, number of signals above the energy threshold and the correlation between the positions of the scintillators in the detector and the time differences between registered signals~\cite{jpet_paper_Kowalski2}.

In the first level of the event selection we defined true coincidence as two Compton scatterings in two different strips with energy deposited larger than the fixed energy threshold. In addition, a number of all scatterings in the event with energy deposition higher than fixed energy threshold must be equal to two and a number of all scatterings with energy deposition higher than a~noise threshold must be two or three (Fig.~\ref{types_of_coincidences}). The noise threshold was set to 10~keV. Moreover, both hits must have fit in time window equal to 3~ns.

The registered signals had to originate from a~single electron-positron annihilation events. In the case of true and detector-scattered coincidences no scattering in the phantom was required. If in addition (i)~both signals with energy larger than the threshold were due to the primary scatterings, the coincidence was marked as a true coincidence (Fig. \ref{types_of_coincidences}a,b), and (ii)~if at least one of the signals above the threshold was due to the secondary scattering, the coincidence was qualified as a detector-scattered coincidence (Fig. \ref{types_of_coincidences}c,d). Otherwise, if at least one of gamma photons was scattered in the phantom before scattering in a~scintillation strip, the coincidence was qualified as a~phantom-scattered coincidence (Fig. \ref{types_of_coincidences}e-h). If hits were caused by gamma photons from different annihilations, the coincidence was marked as an accidental coincidence and rejected.

In order to find selection criteria which would allow a~suppression of the detector-scattered and phantom-scattered coincidences, we have performed studies of the correlation between the detector's identity numbers and the time differences between the registered signals. These investigations are presented in the next section. 



\section{Results and discussion}

\begin{figure}[h!]
\centering

\begin{subfigure}{0.49\textwidth}
\centering
\includegraphics[width=\textwidth]{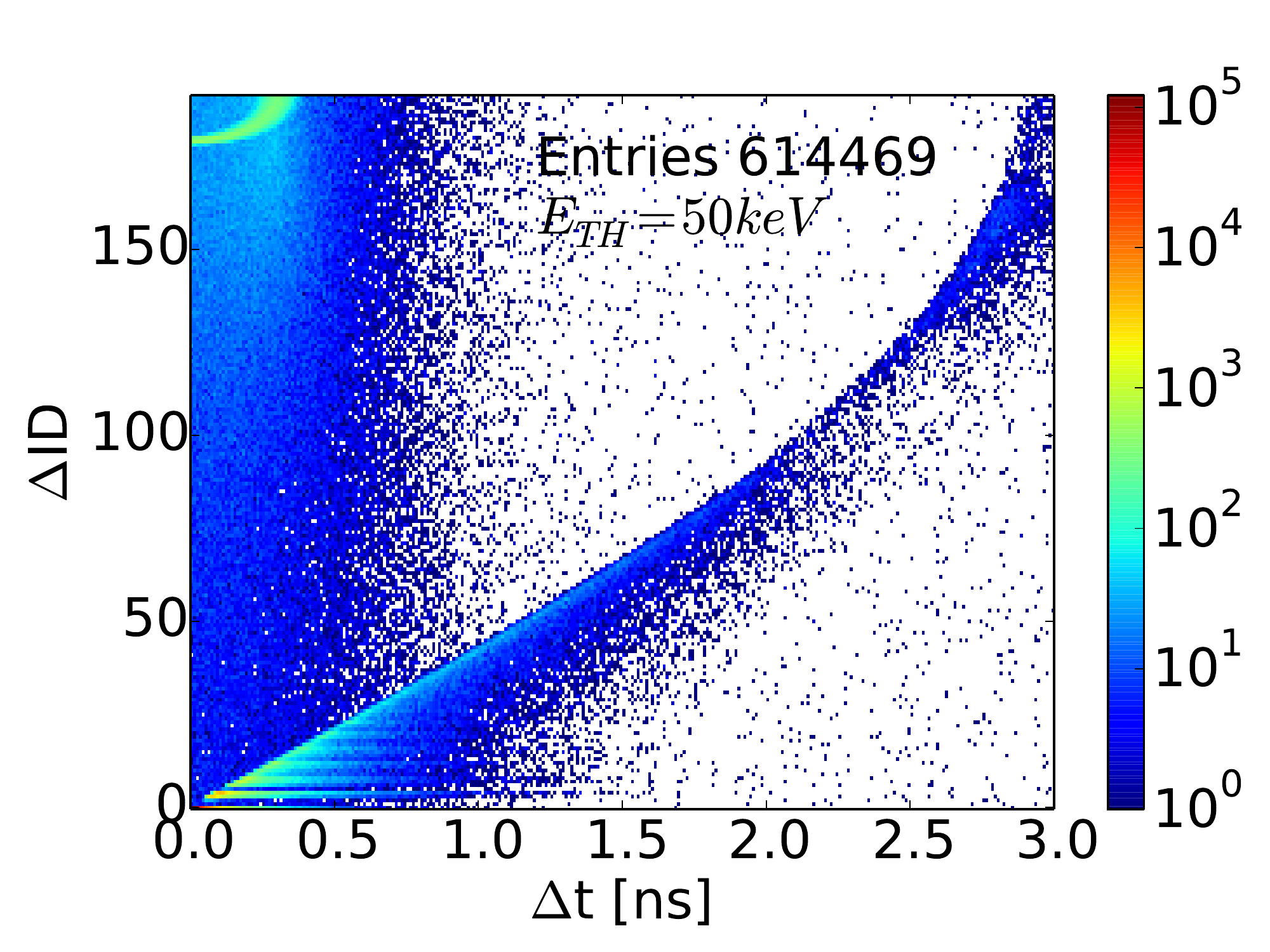}
\end{subfigure}
\begin{subfigure}{0.49\textwidth}
\centering
\includegraphics[width=\textwidth]{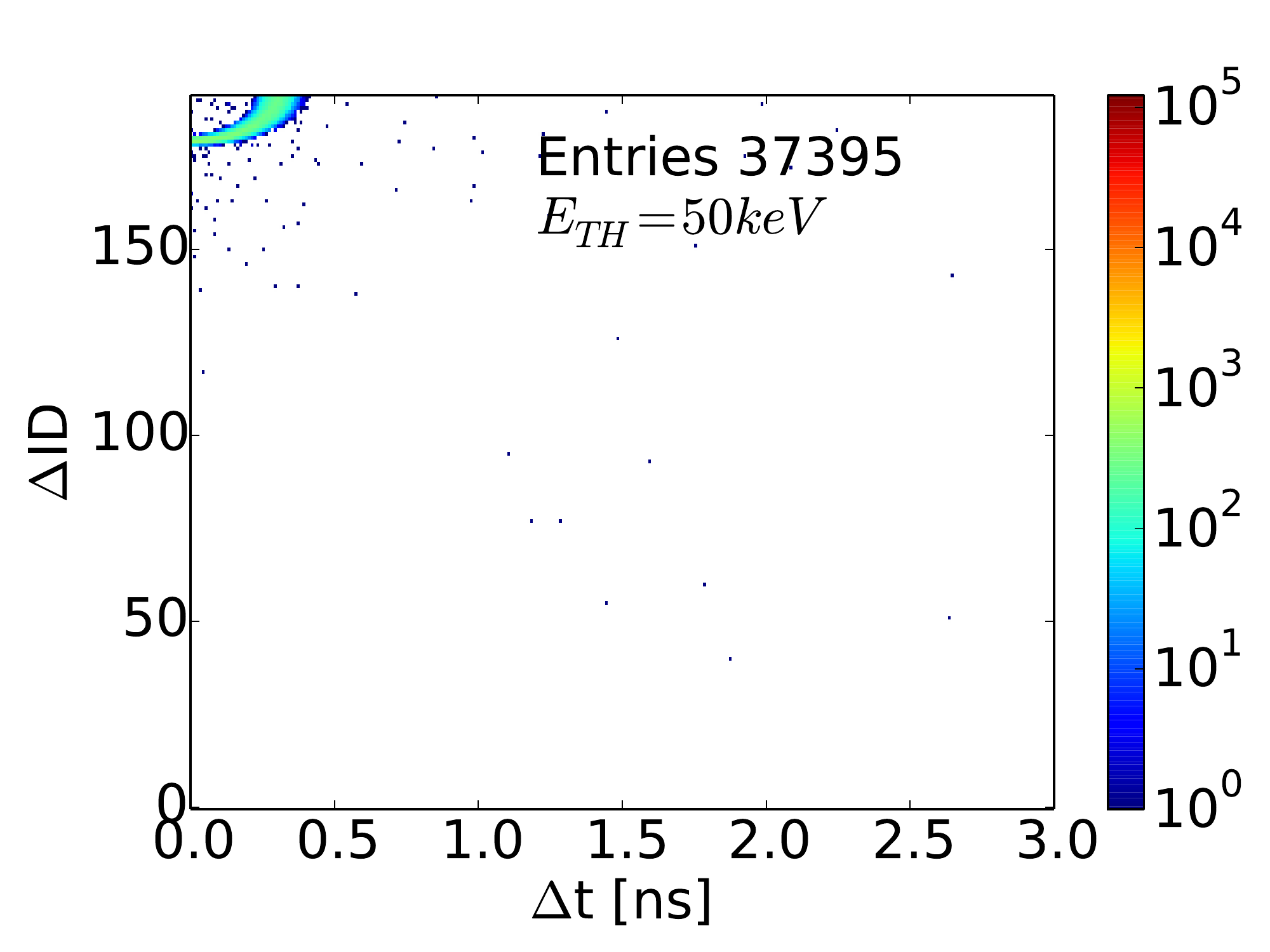}
\end{subfigure}

\begin{subfigure}{0.49\textwidth}
\centering
\includegraphics[width=\textwidth]{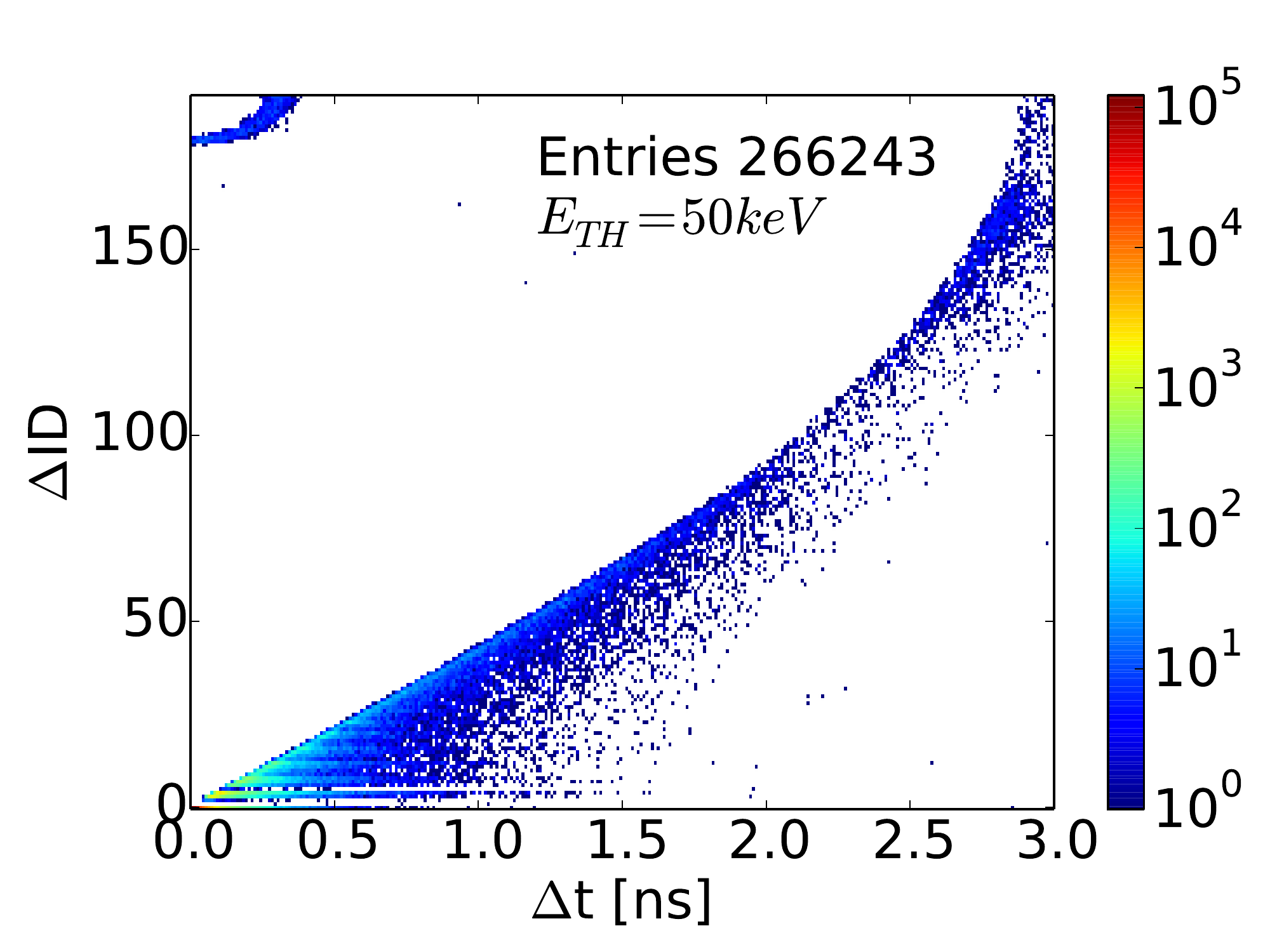}
\end{subfigure}
\begin{subfigure}{0.49\textwidth}
\centering
\includegraphics[width=\textwidth]{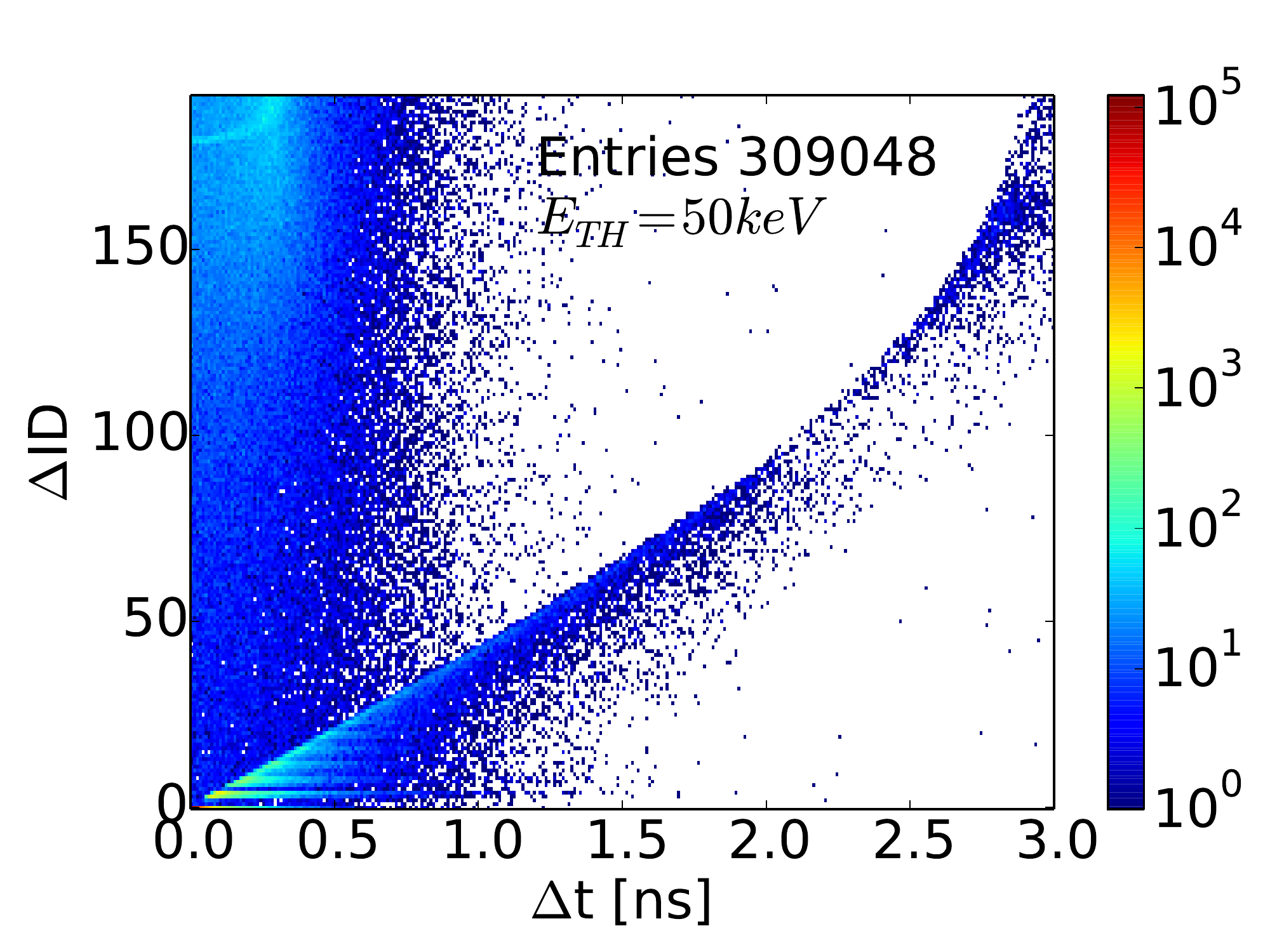}
\end{subfigure}

\caption{Coincidences for the fixed energy threshold equal to 50 keV. All types of coincidences (including accidental ones) are presented in the top left plot. The true, detector-scattered and phantom-scattered coincidences are shown in the top right, bottom left and in the bottom right panels, respectively. $\Delta t$ denotes time difference between subsequent hits and $\Delta\,\mbox{ID}$ denotes the difference between identifiers of scintillator strips (detailed definition in the text).}
\label{Did_vs_Dt_50keV}
\end{figure}

For all coincidences, 2-dimensional histograms of registration time differences between subsequent scatterings $\Delta t$ and scintillator identifiers differences $\Delta\,\mbox{ID}$
were calculated. $\Delta\,\mbox{ID}  = min(|ID_1-ID_2|, 384-|ID_1-ID_2|)$, where $ID_1$ and $ID_2$ denotes numbers of scintillator modules. The scintillator identfiers increase monotonically with the grows of the azimuthal angle and are in the range from 1~to 384~\protect\cite{jpet_paper_Kowalski2}. 
These histograms are presented in Fig. \ref{Did_vs_Dt_50keV}. 
True coincidences are located in the region of low $\Delta t$ and high $\Delta\,\mbox{ID} $.

\begin{figure}[h!]
\centering

\begin{subfigure}{0.49\textwidth}
\centering
\includegraphics[width=\textwidth]{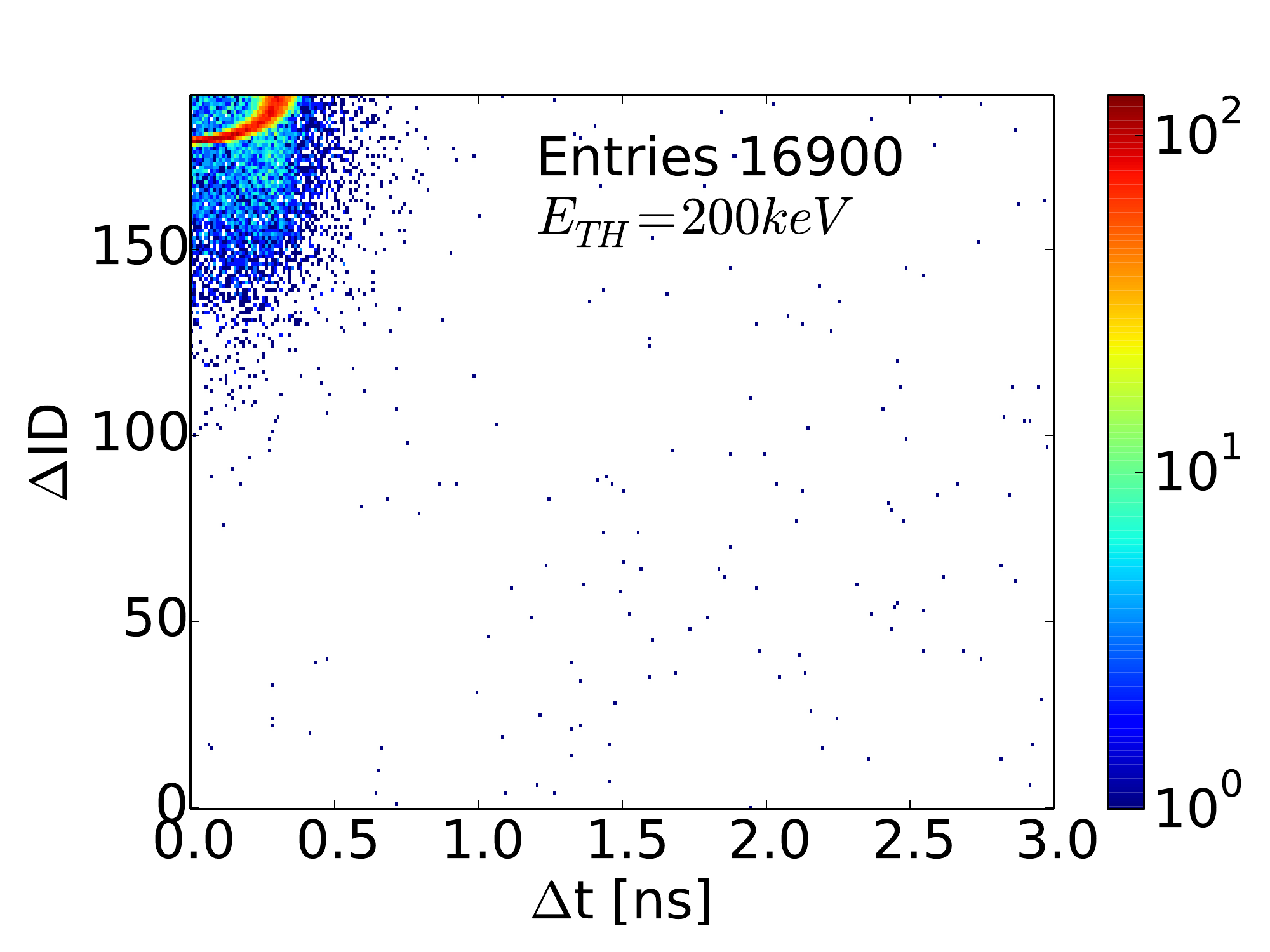}
\end{subfigure}
\begin{subfigure}{0.49\textwidth}
\centering
\includegraphics[width=\textwidth]{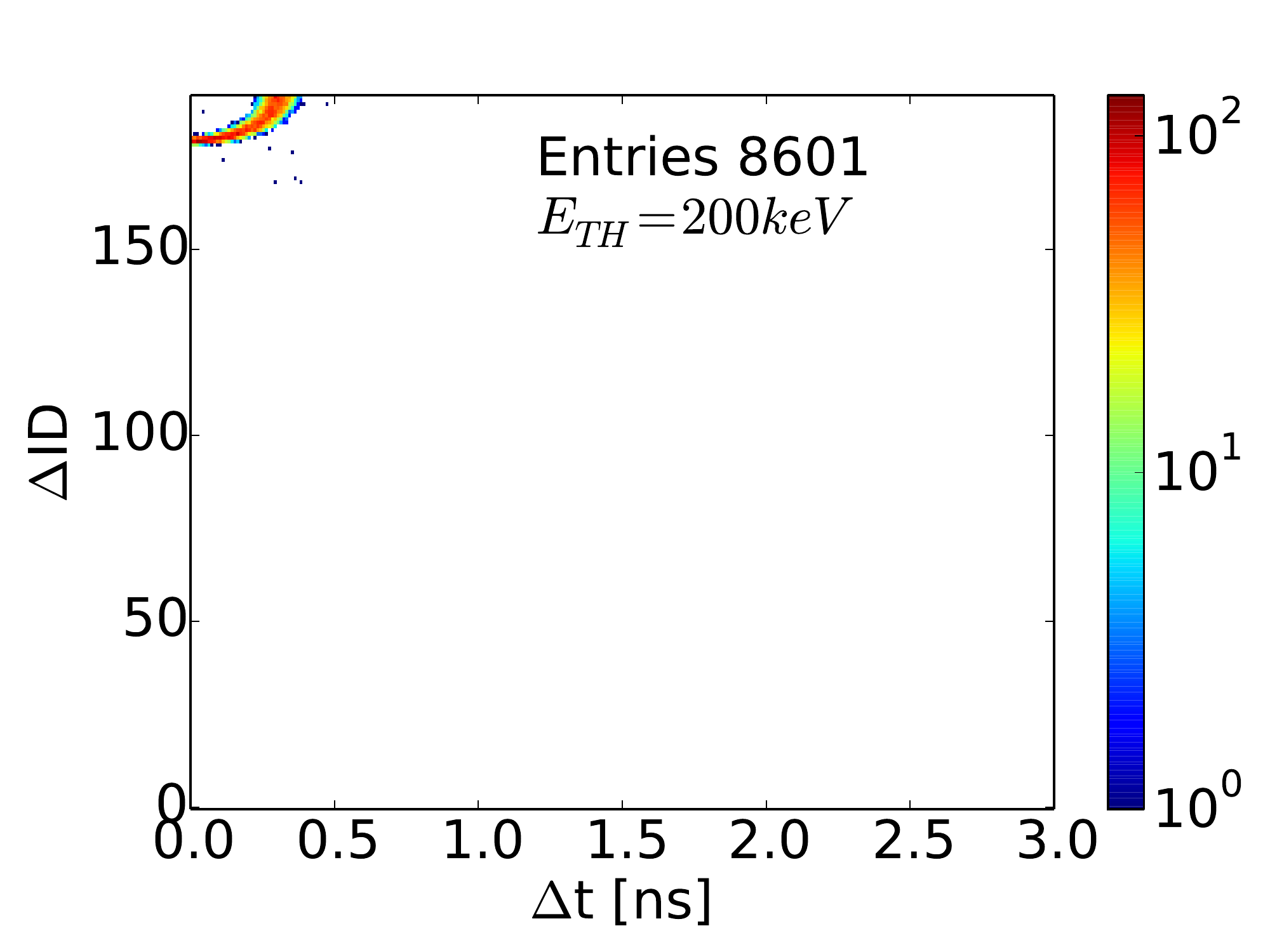}
\end{subfigure}

\begin{subfigure}{0.49\textwidth}
\centering
\includegraphics[width=\textwidth]{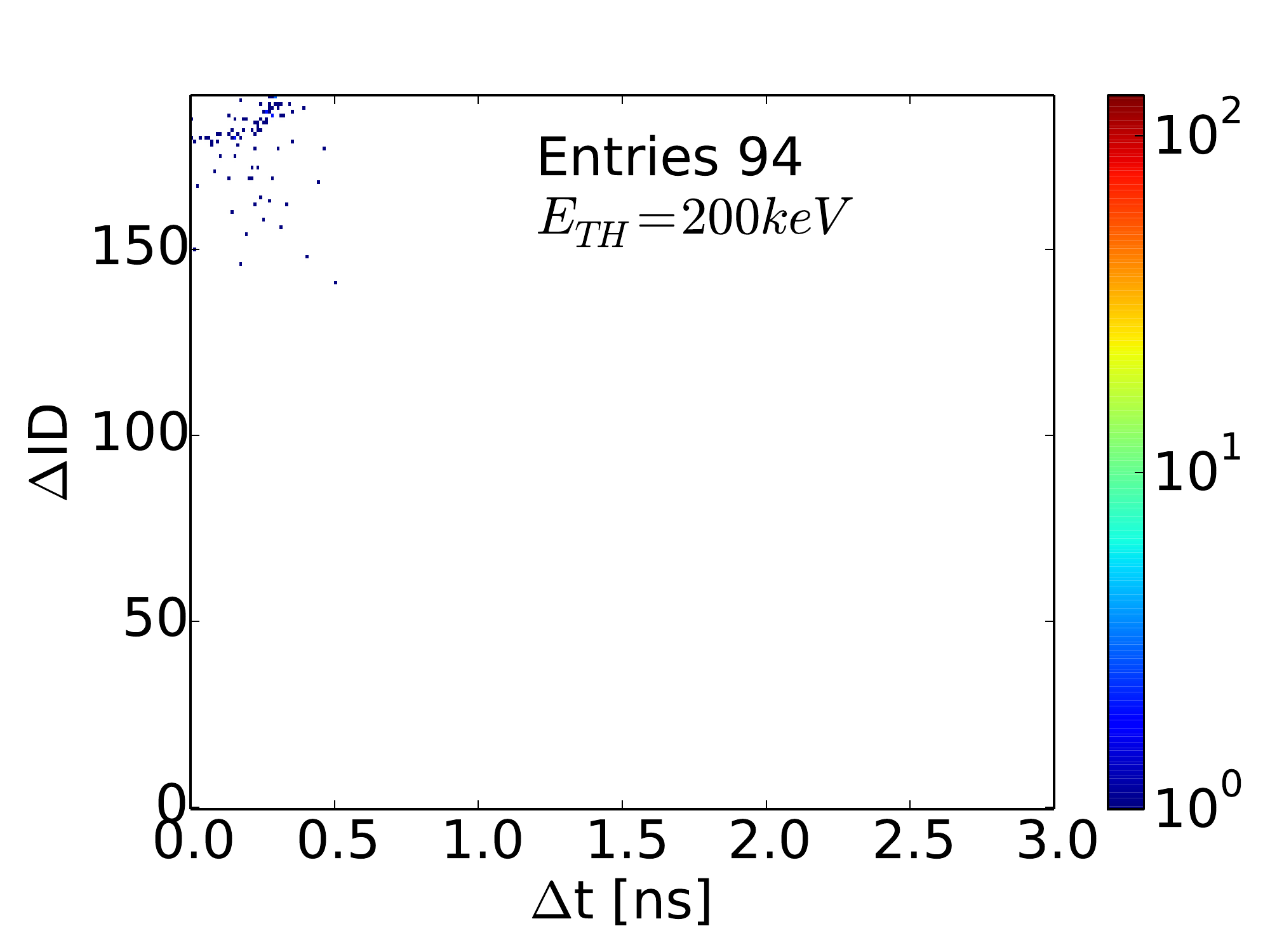}
\end{subfigure}
\begin{subfigure}{0.49\textwidth}
\centering
\includegraphics[width=\textwidth]{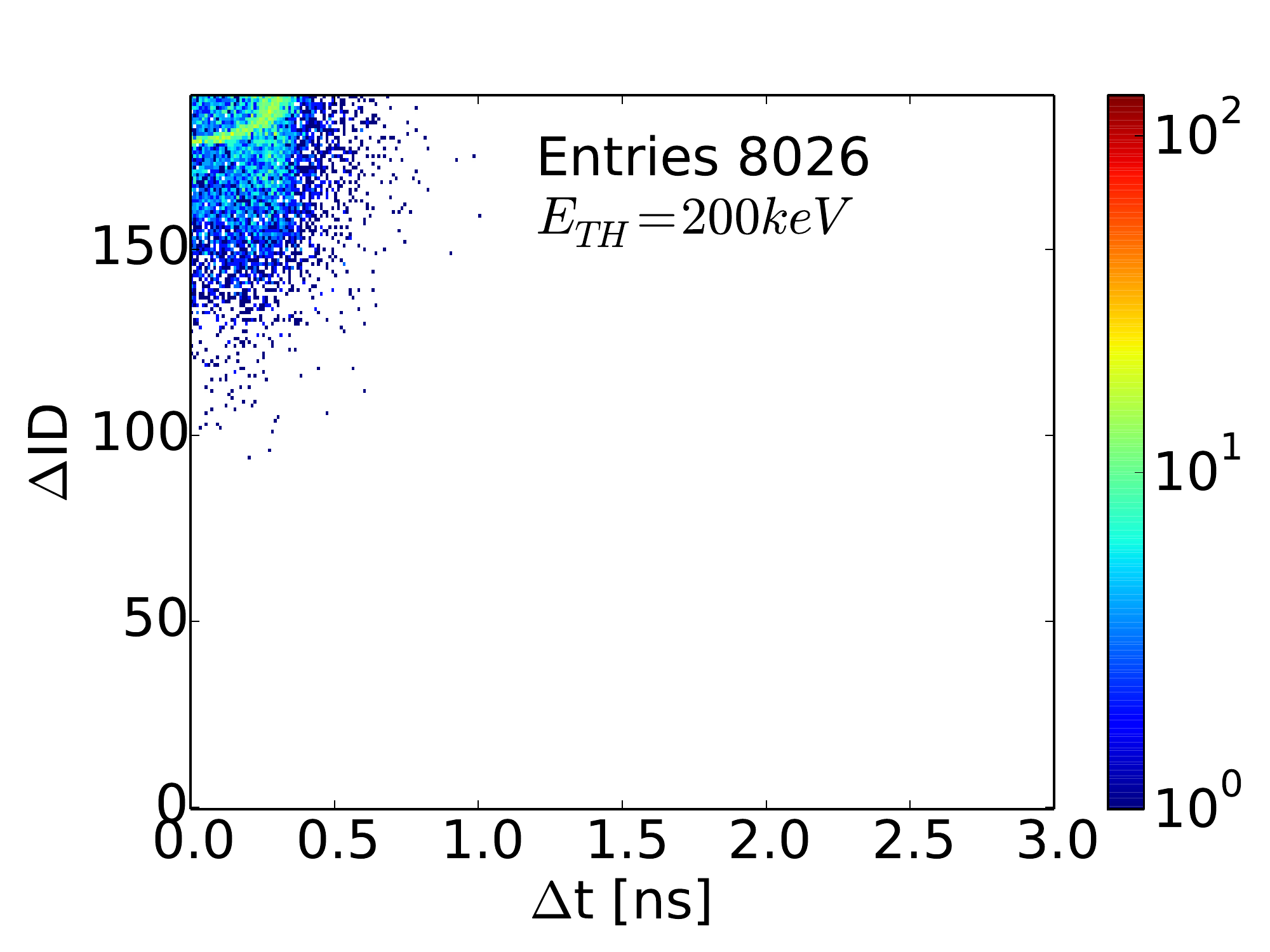}
\end{subfigure}

\caption{Coincidences for the fixed energy threshold equal to 200 keV. All types of coincidences (including accidental ones) are presented in the top left plot. The true, detector-scattered and phantom-scattered coincidences are shown in the top right, bottom left and in the bottom right panels, respectively. $\Delta t$ denotes time difference between subsequent hits and $\Delta\,\mbox{ID}$ denotes the difference between identifiers of scintillator strips (detailed definition in the text).}
\label{Did_vs_Dt_200keV}
\end{figure}

If one applies energy threshold equal to 200~keV, the ratio between the number of true and scattered coincidences increases by a~factor of about 16 as it is shown in the Fig.~\ref{Did_vs_Dt_200keV}.
The number of scattered coincidences is reduced about 70 times while the number of true coincidences is smaller by about 4 times. 

\begin{figure}[h!]
\centering

\begin{subfigure}{0.49\textwidth}
\centering
\includegraphics[width=\textwidth]{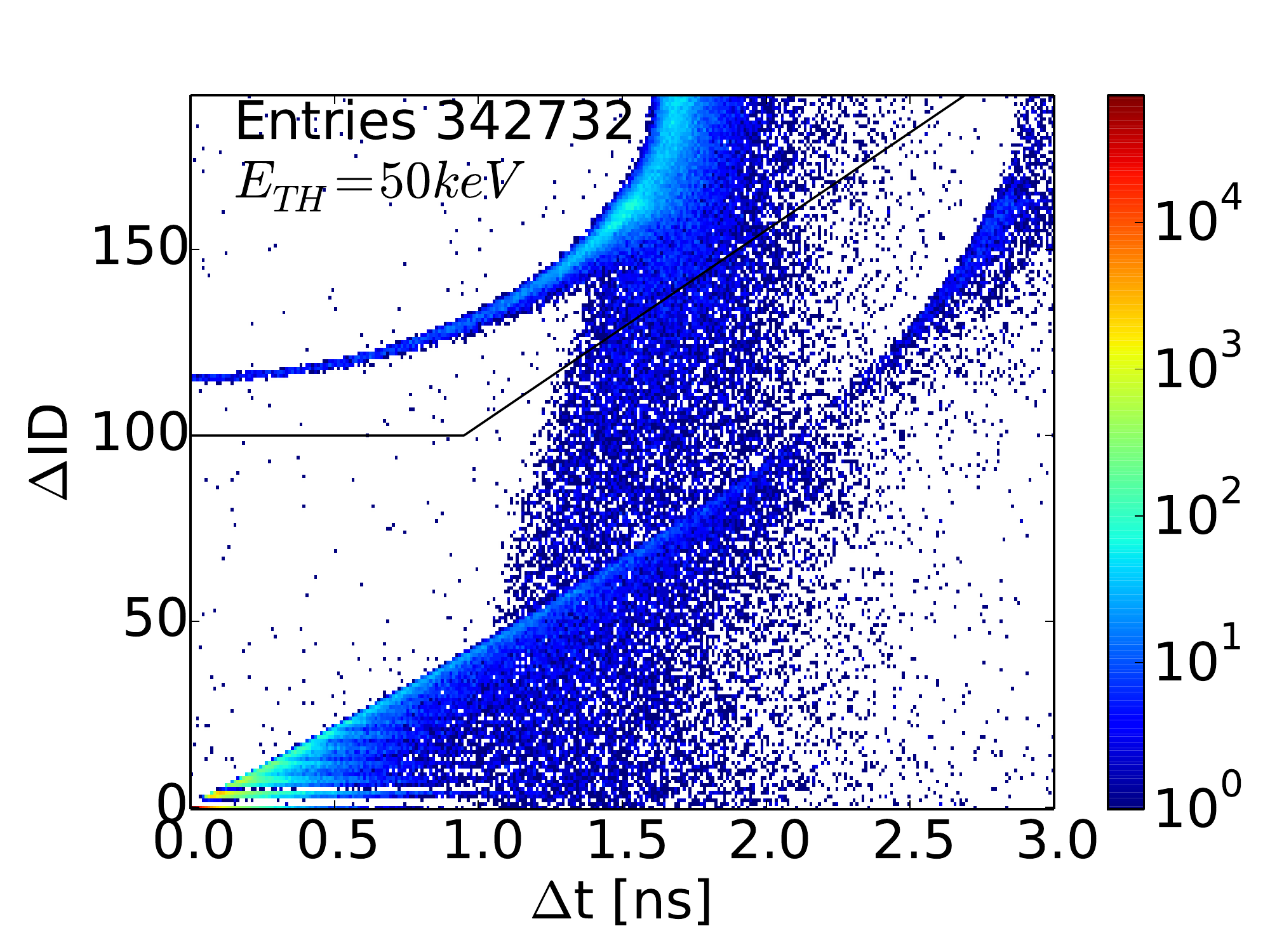}
\end{subfigure}
\begin{subfigure}{0.49\textwidth}
\centering
\includegraphics[width=\textwidth]{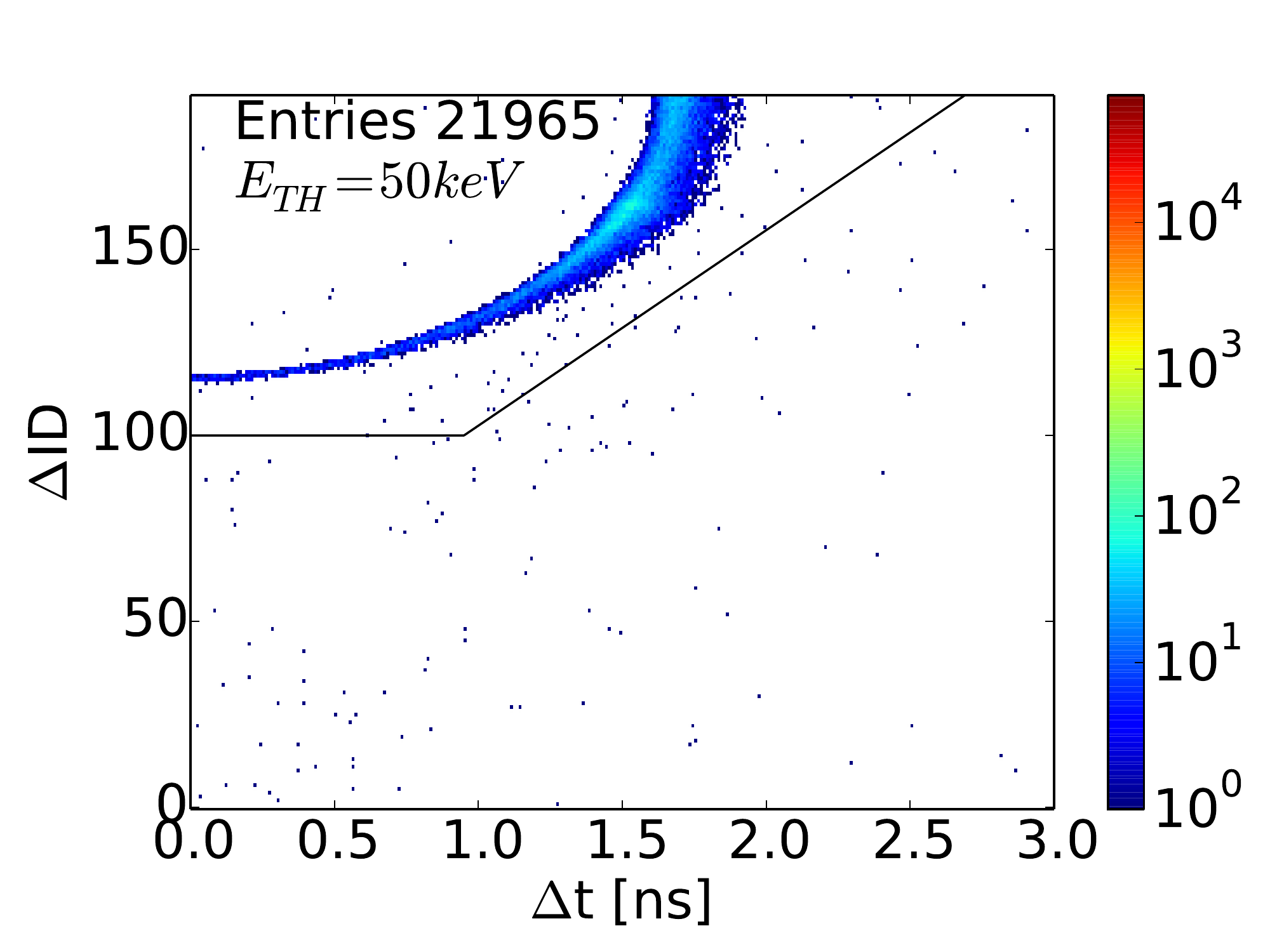}
\end{subfigure}

\begin{subfigure}{0.49\textwidth}
\centering
\includegraphics[width=\textwidth]{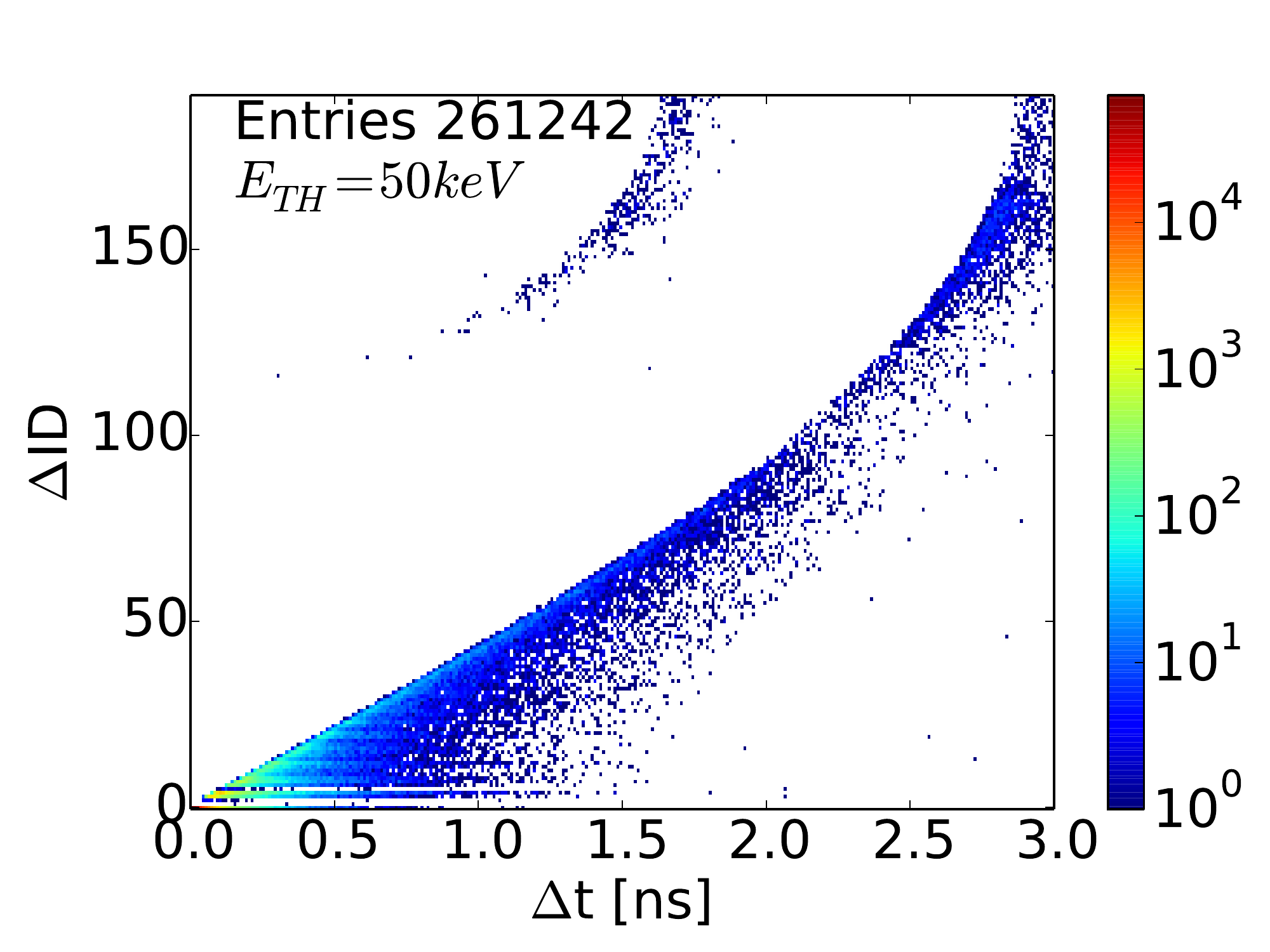}
\end{subfigure}
\begin{subfigure}{0.49\textwidth}
\centering
\includegraphics[width=\textwidth]{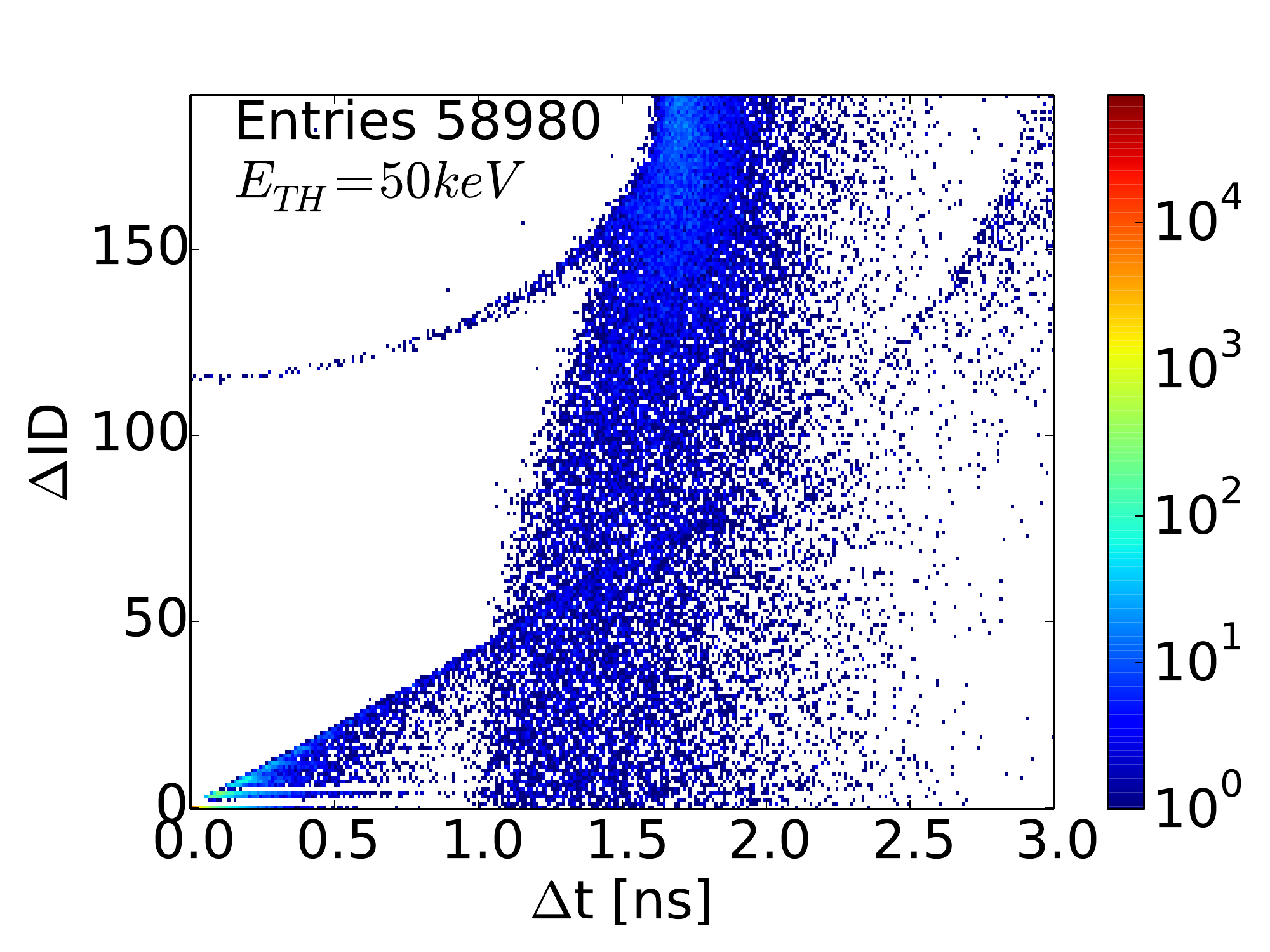}
\end{subfigure}

\caption{$\Delta\,\mbox{ID} $ vs $\Delta t$ correlations for evens simulated with the source moved to the radial distance of 25 cm. In the data selection a~50~keV energy threshold was applied.
The black line is a visual representation of the second level selection: $\Delta\,\mbox{ID}$~$<$~$100$ and point is below the line connecting points (0~ns, 50) and (2.7~ns, 192). True coincidences lie above this line.}
\label{Did_vs_Dt_50keV_source25cm}
\end{figure}

In order to check how to minimize the number of scattered coincidences by using correlations of $\Delta t$ vs $\Delta\,\mbox{ID} $, additional simulations were performed. The linear source was placed outside the scatter phantom at the radial distance of 25~cm. Results of such simulations may be used to estimate scattering rejection criteria for big objects with diameters of about 50~cm (for example human body). These results are presented in Fig. \ref{Did_vs_Dt_50keV_source25cm}. As one can see, most of detector-scattered coincidences we could reject if we took into account only coincidences with $\Delta\,\mbox{ID} $ difference higher than 100 (which corresponds to the azimuthal angle distance equal to about 94 degrees) and lying above the line connecting points (0~ns,~50) and (2.7~ns,~192).
For the final selection, in addition to the criteria described in the section \ref{methods}, we apply the graphical $\Delta\,\mbox{ID} $ vs $\Delta t$ condition shown in Fig. \ref{Did_vs_Dt_50keV_source25cm}.   

In Fig. \ref{Did_vs_Dt_50keV_source25cm}, for detector-scattered coincidences, there is longitudinal structure extending between points (0 ns, 0) and (3 ns, 192). These events correspond to differences between times of primary photon reactions in a~given scintillators and times of the secondary interactions. The larger is the angle of the primary scattered photon, the larger will be the $\Delta\,\mbox{ID}$ and $\Delta t$. E.g. the bin with coordinates (2.9 ns, 192) corresponds to the backscattering - primary particle is backscattered and it is registered in the strip on the opposite side of the detector (2.9 ns is the time needed by the photon to travel between opposite strips with speed of light in vacuum) \cite{jpet_paper_Kowalski2}.
In the case of the phantom-scattered coincidences, the $\Delta\,\mbox{ID}$ may vary from zero to 192 as can be inferred from Figs. \ref{types_of_coincidences}e,f. Maximum value of time difference $\Delta t$ corresponds to the time of flight of the gamma photon along the diameter of the scanner, which is equal to about 2.9~ns.

\begin{figure}[h!]
\centering

\begin{subfigure}{0.49\textwidth}
\centering
\includegraphics[width=\textwidth]{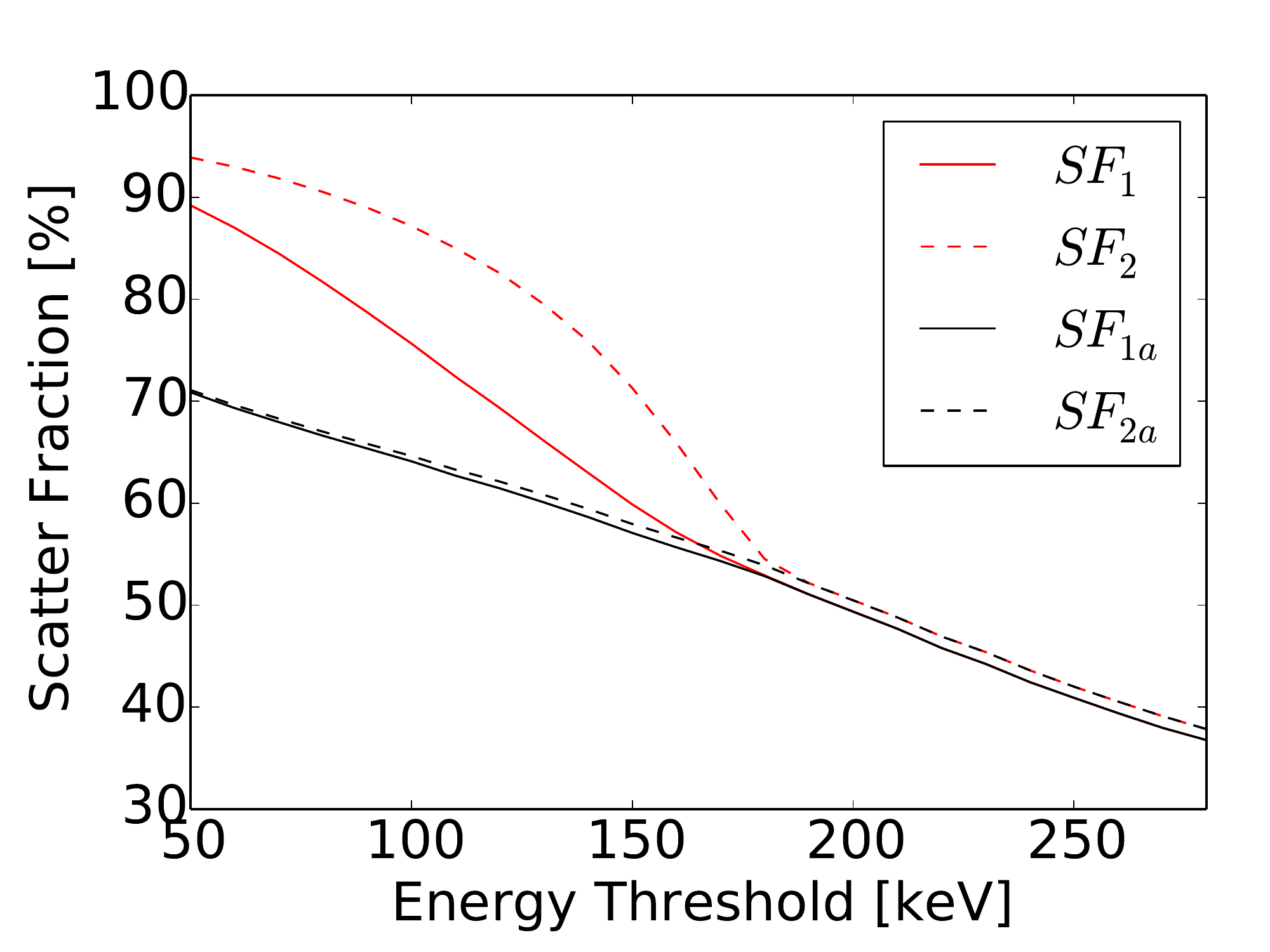}
\end{subfigure}
\begin{subfigure}{0.49\textwidth}
\centering
\includegraphics[width=\textwidth]{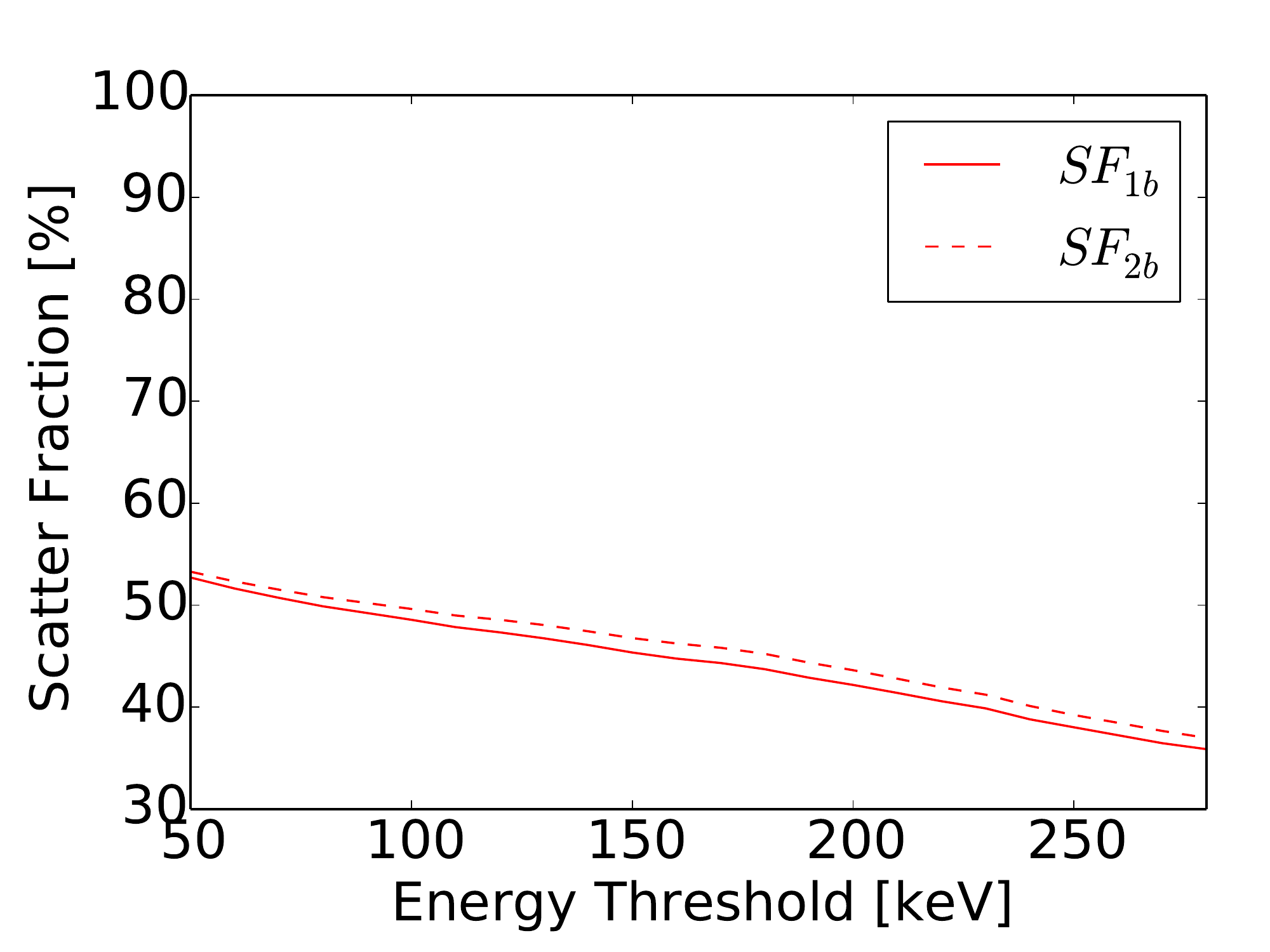}
\end{subfigure}

\caption{Scatter fraction of the 384-strip J-PET detector as a~function of energy threshold calculated using uncut data from GATE simulation. The left panel shows the dependencies of scattered fraction calculated after the first level selection (lines $SF_1$ and $SF_2$) and after the second level selection using a~$\Delta\,\mbox{ID} $ vs $\Delta t$ correlation (lines $SF_{1a}$ and $SF_{2a}$). 
The SF determined based on the NEMA norm (lines $SF_{1b}$ and $SF_{2b}$) are shown in the right panel.
More detailed description in the text. The result shown is based on $10^8$ simulated annihilations.}
\label{scatter_fraction}
\end{figure}

Dependencies of SF on energy threshold for events fulfilling the first and second level of selection criteria, are presented in Fig. \ref{scatter_fraction}.
As one can see in the left panel, after the first-level selection, value of scatter fraction calculated using only true and phantom-scattered coincidences (red dotted line $SF_{1}$) depends approximately linearly on the energy threshold. The scatter fraction calculated using true and both phantom- and detector-scattered coincidences (red continuous line $SF_{2}$), collapses at the energy threshold equal to about 180 keV. This effect will be explained in the next paragraphs. 

The most probable detector-scattered coincidence is a~coincidence in which there is one primary and one secondary scattering. In order to have such a~scenario, energy threshold must be smaller than certain physical limit. This limit is caused by two contradictory conditions.

On the one hand, the energy deposited in the primary scattering of gamma photon (with energy $E_1$ = 511 keV) should be larger than $E_{TH}$ (energy threshold), which corresponds to the condition that the energy of the scattered gamma photon, after primary scattering must be smaller then the given energy value $E_{S1}$ (Eq.~\ref{eq:Es1}~and~\ref{eq:thetamin}).

On the other hand, in order to have a~possibility that the second energy deposition is bigger than $E_{TH}$, energy of scattered photon must be higher than $E_{S2}$ (Eq.~\ref{eq:Es2}). These two contradictory conditions give system of two equations. From these equations, one could obtain value of $E_{TH}$ which satisfies both of them, which is 184~keV. If energy threshold would be larger than 184~keV, there should be no detector-scattered coincidences.

\begin{equation} 
\label{eq:Es1}
E_{S1} = {  \left( {1 \over E_1} + {1 \over {m_e c^2}} (1 - \cos \theta_{min}) \right)^{-1}  }
\end{equation}

\begin{equation} 
\label{eq:thetamin}
\theta_{min} = {\arccos {(1 - {{m_e c^2 E_{TH}} \over {E_1 (E_1 - E_{TH})}}}} )
\end{equation}

\begin{equation} 
\label{eq:Es2}
E_{S2} = 0.5 ( E_{TH} + \sqrt{ E_{TH} (E_{TH} + 2 m_e c^2)})
\end{equation}

SF strongly depends on event selection using $\Delta\,\mbox{ID} $ vs $\Delta t$ correlation plot. The application of first- and second- level selection criteria is presented as black lines in Fig. \ref{scatter_fraction}. The biggest change is visible in the range of low energy thresholds. The most of detector-scattered coincidences are eliminated and the same trend for phantom-scattered coincidences can be observed.


The NEMA-NU-2 norm \cite{norm_nema} defines the criteria that describes to the size of the scatter phantom.
According to these criteria, all pixels located further than 12 cm from the centre of the transaxial field of view of the scanner shall be set to zero. If they would be used for $\Delta\,\mbox{ID} $ vs $\Delta t$ correlation plot, one could obtain more visible reduction of scatter fraction equal at most 50\% for very low energy threshold equal to 50~keV. 
For threshold equal to about 200~keV, value of scatter fraction would be even 35\%.
Dependencies of scatter fraction on the fixed energy threshold fot NEMA-based criteria are presented in the right panel of Fig.~\ref{scatter_fraction}.


\subsection{Summary and comparison to commercial scanners}

SF obtained for the single-layer J-PET scanner varies from 37\% to about 70\% depending on the value of fixed energy threshold and for the classical definition of the scatter fraction including only scattering in the phantom. E.g. for the fixed energy threshold equal to 200 keV, SF varies from 42 \% to about 50 \% depending o the definition used and on additional criteria. If one applies criteria based on NEMA-NU-2 norm, SF is in the range from 37\% to 53\% depending on the value of the fixed energy threshold.
If detector-scattered coincidences were taken into account, then SF  would be larger in the region from 50~keV to about 184~keV. Above the value, no detector-scattered coincidence can appear.

Scatter fraction for PET scanners is on the same order of magnitude as for the J-PET scanner.
In the RPC-PET scanner, which is also a~large axial field of view scanner and is in some ways similar to J-PET scanner, value of scatter fraction varies from 50\% to 60\% \cite{sf_reference_rpc}.
For most of commercial PET scanners, scatter fraction is between 30\% and 40\% \cite{sf_reference_generally}. 
Looking into more detail, example values of SF for chosen models of known producers are: about 40\% for GE Discovery PET/CT scanner \cite{sf_reference_ge_discovery}, 27-45\% for Philips Gemini TF PET/CT scanner (depending on the diameter of the phantom from 20~cm to 35~cm) \cite{sf_reference_philips_gemini} and about 31-34\% for Siemens BiographTM~6\cite{sf_reference_biographTM6}. Values of SF for different models of PET scanners are summarized in Tab.~\ref{table_comparison}.

\begin{table}[t]
\centering
\caption{Comparison of SF for chosen models of PET scanners}
\label{table_comparison}
\begin{tabular}{|c|c|}
  \hline 
  \textbf{PET scanner} & \textbf{SF [\%]}\\
  \hline
  RPC-PET & from 50 to 60 \\
  \hline
  GE Discovery PET/CT & about 40 \\
  \hline
  Philips Gemini & from 27 to 45 \\
  \hline
  Siemens Biograph TM & from 31 to 34 \\
  \hline
  J-PET & from 37 to 53 \\
  \hline
\end{tabular} 
\end{table}

\section{Acknowledgements}
We acknowledge technical and administrative support by  A. Heczko, M. Kajetanowicz, W. Migda\l, and the financial support by The Polish National Center for Research and Development through grant No. INNOTECH-K1/IN1/64/159174/NCBR/12, and through LIDER grant 274/L-6/14/NCBR/2015, and The Foundation for Polish Science through MPD program and the EU, MSHE Grant No. POIG.02.03.00-161 00-013/09, and Marian Smoluchowski Krakow Research Consortium "Matter-Energy-Future".

\end{document}